\documentclass[pra,twocolumn,floatfix]{revtex4}
\usepackage{times,amsmath,amssymb,amstext,latexsym,float,graphicx,color}
\usepackage[hidelinks]{hyperref}
\hypersetup{colorlinks=true, citecolor=blue, urlcolor=blue, linkcolor=blue}
\usepackage[normalem]{ulem} 
\usepackage{cancel}

\begin{document}

\title{Dipolar particles in a double-trap confinement: Response to tilting the dipolar orientation}

\author{J. Bjerlin$^1$, J. Bengtsson$^1$, F. Deuretzbacher$^2$, L.H. Kristinsd\'ottir$^1$  and S. M. Reimann$^1$}
\affiliation{$^1$Mathematical Physics and NanoLund, LTH, Lund University, P.O.Box 118, SE-22100 Lund, Sweden\\
$^2$Institut f\"ur Theoretische Physik, Leibniz Universit\"at Hannover, Appelstrasse 2, DE-30167 Hannover, Germany }
\date{\today}

\begin{abstract}
We analyze the microscopic few-body properties of dipolar 
particles confined in two parallel quasi-one-dimensional harmonic
traps.  In particular, we show that an adiabatic  rotation of the dipole orientation  about the trap axes can  drive an initially non-localized few-fermion state into a localized state with strong inter-trap pairing. 
For an instant, non-adiabatic  rotation, however,  localization is inhibited and a highly excited state is reached. This state may be interpreted as the few-body analog of a super-Tonks-Girardeau state, known from one-dimensional systems with contact interactions.  
\end{abstract}

\maketitle 

\section{Introduction}
Ultra-cold atoms or molecules with large permanent magnetic~\cite{griesmaier2005,lahaye2007,koch2008,beaufils2008,lu2010, lu2011,aikawa2012,Lu2012,aikawa2014}, or electric~\cite{aymar2005,ni2008,deiglmayr2009,aikawa2010,danzl2010,takekoshi2014}, dipole moments have  interesting properties that originate from the spatial anisotropy of the dipole-dipole interaction (DDI). 
 Also the trap geometry  is, in general, decisive for the properties of these ultra-cold systems. Different quasi one- and two-dimensional (1D and 2D) 
confinement configurations,  ranging from single-traps to optical lattices and coupled interlayer systems,  have been considered,  see, e.g., Refs.~\cite{baranov2008,lahaye2009} for reviews. 
The reduced dimensionality, together with the fact that the DDI can be relatively strong and tailored by means of external fields~\cite{sinha2007cold,buchler2007} makes it possible to study  
strongly correlated systems.
The long-range part of the dipolar interaction may couple spatially distinct confinement regions in different ways, leading to intriguing many-body phenomena in optical lattices and multilayer systems, see, for example, Refs.~\cite{Goral2002, chang2009, huang2009, zinner2011, chotia2012, macia2014single,pizzardo2016,hebenstreit2016dipolar}.  
In the few-body regime,  bound states of dipoles have been predicted~\cite{wunsch2011,dalmonte2011trimer,knap2012,volosniev2013bound}.   
For instance, in a planar array of parallel 1D tubes, inter-tube interactions were found responsible for the formation of clustered states of Wigner- or 
Luttinger type~\cite{knap2012}. 

In 1D, a  strong contact repulsion may lead to a so-called Tonks-Girardeau (TG) state, where bosons show properties resembling non-interacting 
fermions~\cite{girardeau1960relationship}. The cold-atom realization of such states~\cite{kinoshita2004observation, paredes2004tonks}  spurred further work to probe their characteristic properties~\cite{astrakharchik2004quasi,astrakharchik2005beyond,deuretzbacher2007evolution,
tempfli2008excitations,murmann2015antiferromagnetic}.
States  strongly resembling the TG ones  have, in addition, been seen for systems of dipolar bosons~\cite{deuretzbacher2010, girardeau2012super}. Also,
TG states are closely related to the excited states of 1D-confined particles with short-range {\it attractive} interactions, 
known as super-Tonks-Girardeau (STG) states~\cite{astrakharchik2005beyond, chen2010transition,girardeau2012super}.   
For a more detailed theoretical description and experimental realization of the STG states, see, for example, Refs.~\cite{astrakharchik2004quasi,astrakharchik2005beyond,tempfli2008excitations,haller2009realization,wunsch2011,girardeau2012super,chen2010transition}. 
A system initially in a TG state may, for instance with a sudden quench of the  interaction, end up in a STG state \cite{girardeau2012super}. 

Here, we show that in parallel 1D traps, depending on the dipole angle, the DDI  may lead to fundamentally different few-body ground states. The interplay between the particle interactions \textit{within} each trap and \textit{between} the traps plays a decisive role in the formation of a localized ground state in each  wire.   
We also investigate the transition from a TG- to a STG-like state by a sudden quench of the dipole angle, and find that it is possible to achieve this transition even 
for spin-polarized fermions. 
Our work is partly motivated by the common use of a contact pseudopotential to approximate the DDI~\cite{volosniev2013bound}, which has been utilized to, for example, study the phase diagram of two parallel 1D traps~\cite{huang2009quantum,lecheminant2012exotic,fellows2011superfluid}. We here consider the full form of the DDI, and specifically address emergent short-  {\it and} long-range phenomena in the exact solutions of the few-body system.

\section{Model}
\label{sec:model}
\begin{figure}[ht]
\includegraphics[width=0.8\linewidth]{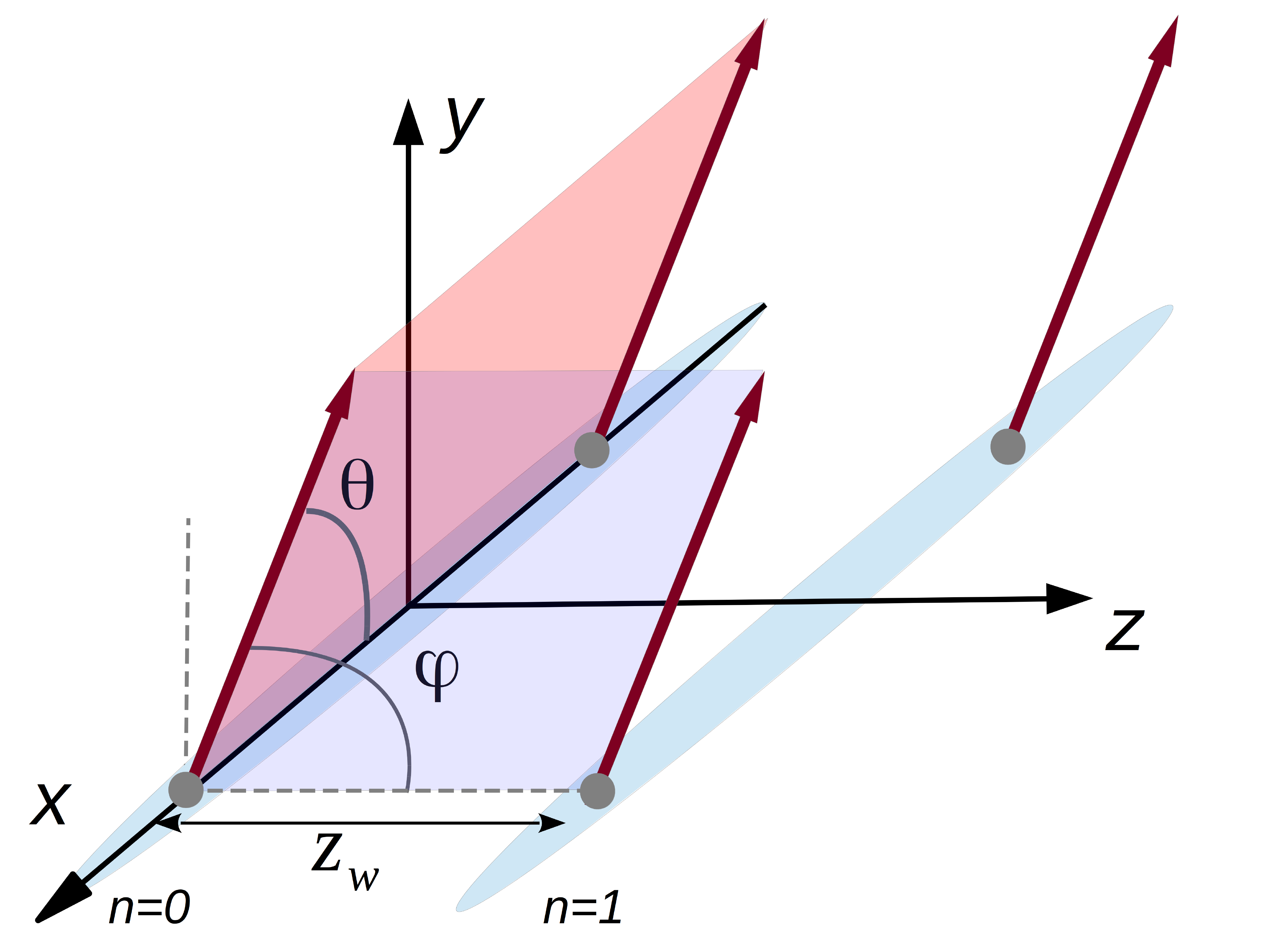}
\caption{(Color online) Two parallel traps, sketched by the light-blue areas, elongated along the $x$-direction and with $z_w$ being the shortest distance between them. Each trap
  contains two dipolar particles illustrated by the red bullets with connecting arrows.  The dipoles are aligned
  with an external field. The angles $\phi$ and $\theta $
  define the direction of alignment relative to the $z$- and $x$-axis,
  respectively.  } \label{Fig1}
\end{figure}

Let us consider  (magnetic or electric) dipolar
particles in two parallel, cigar-shaped traps  or wires,  as schematically 
sketched in Fig.~\ref{Fig1},
\begin{equation}
V^{\text{trap}}_n(\mathbf{r}) =  \frac{m}{2} \omega_x^{2} x^{2} +
\frac{m}{2}\omega_{\bot}^{2}\left[ y^2+\left(z-nz_w\right)^{2}
  \right] \label{Vtrap}~.
\end{equation} 
In Eq.~\ref{Vtrap}, the index $n=\{0,1\}$  labels 
the two traps, $m$ is the particle mass, ${\bf
  r}=(x,y,z)$ and the oscillator frequencies satisfy 
$\omega_x\ll\omega_\bot$.
The strong confinement perpendicular to the wire axes is here sufficient to neglect the tunneling of particles between the two traps, which keeps the number of particles per trap constant. 
Furthermore, $\omega_\bot$ is assumed strong enough such that the 
dipole interaction is insufficient to excite the system in the tightly confined directions, which then can be adequately described by the harmonic
oscillator ground state. For the particles trapped by $V^{\text{trap}}_n$ 
we may then  consider a  basis 
\begin{equation}
\psi_{n,j}(\mathbf{ r}) =  \varphi_j(x) \phi_{n}(y,z), \label{one-body orb}
\end{equation}
where $\varphi_j$ are the basis functions in the
$x$-direction (as further specified in the appendix), and
\begin{equation}
 \phi_{n}(y,z) = \frac{1}{l_\bot \sqrt{\pi}}
e^{-[y^2+(z-nz_w)^2]/2l_\bot^2} \label{hogr}
\end{equation} 
is the ground-state wave function of the two-dimensional harmonic
oscillator centered at $(y,z)=(0,nz_w)$. In Eq.~(\ref{hogr}), $l_\bot
= \sqrt{\hbar/(m\omega_\bot)}$ is the characteristic length of the
$yz$-oscillator. This length needs to be
shorter than the distance between the two traps to ensure 
a clear separation of the two confinements with $n=0$ and $n=1$. 
The restrictions in the $yz$-plane conveniently allow a quasi 1D description of the double-trap system.

We now address the dipolar interaction between two particles. 
The angle between the particles' dipolar vectors 
determines the anisotropy of the interaction. 
For identical dipole moment vectors
${\bf d}$, the interaction potential reads (see, for example, 
Ref.~\cite{jackson1999classical})  
\begin{equation}
V\left(\tilde{\mathbf{r}}\right)=\frac{\mathbf{d}^{2}}{4\pi \gamma
  \left| \tilde{{\bf
      r}}\right|^{3}}\left[1-\frac{3\left(\mathbf{d}\cdot\tilde{\mathbf{r}}\right)^{2}}{{\bf
      d}^{2} {\tilde{\bf r}} ^{2}}\right], \label{VDDI}
\end{equation}
where $\tilde{{\bf r}}={\bf r} - {\bf r}^{\prime}$ is the relative
position of the two dipolar particles. In the case of electric dipoles, 
$\gamma = \epsilon_0$, and for
magnetic dipoles, $\gamma = 1/\mu_0$, where $\epsilon _0$ and $\mu_0$ are the permittivity and permeability of free space, respectively. 
In the quasi-1D description of the double-trap system, we
use an effective potential, $V^{\text{\it eff}}_{n,n^\prime}$, to model
the interaction. The indices $n$ and $n^\prime$ discern the two different effective potentials; one {\it intra-trap}
potential for interacting particles in the same trap ($n=n^{\prime}$)
and one {\it inter-trap} potential for particles in opposite traps ($n
\neq n^{\prime}$). Both potentials are derived from the integral
expression
\begin{align}
& V^{\text{\it eff}}_{n,n^{\prime}}(\tilde{x})= \nonumber \\ & \iiiint
  dydzdy^{\prime}dz^{\prime} \left| \phi_{n}(y,z)\right|^2 \left|
  \phi_{n^{\prime}}(y^{\prime},z^{\prime}) \right|^2 V(\tilde{{\bf
      r}}),\label{VDDIeff}
\end{align}
where  $\tilde{x}=x-x^{\prime}$. 

For two particles in the {\it same} trap,
Eq.~(\ref{VDDIeff}) reduces to
\begin{align}
& V^{\text{\it eff}}_{n,n}(\tilde{x}) = \frac{{\bf d}^2\left[1+3\cos(2\theta)\right]}{16\pi \gamma l^3_\bot}  \nonumber \\ & \times \left[\frac{\lvert \tilde{x}\rvert}{l_\bot}
-  \sqrt{\frac{\pi}{2}} \left(1+\frac{\tilde{x}^2}{l^2_\bot}\right) \text{erfc}\left(\frac{\lvert\tilde{x}\rvert}{\sqrt{2}l_\bot} \right) e^{\tilde{x}^2/2l^2_\bot} \right. \nonumber \\   & \left.+ \frac{4 l_\bot}{3}\delta(\tilde{x})\right],
\label{VDDIintra}
\end{align}
as, e.g., demonstrated by Sinha and Santos~\cite{sinha2007cold}. We repeat that in Eq.~(\ref{VDDIintra}), $\theta$ is
the angle between the dipole vectors and the $x$-axis, as indicated in
Fig.~\ref{Fig1}. The intra-trap potential is clearly independent of
the angle $\phi$ between the dipole vectors and the $z$-axis. Hence,
depending on $\theta$ alone, the interaction can be either attractive,
zero or repulsive. In particular, the potential is maximally repulsive
for $\theta = 90^\circ$ and vanishes at the ``critical angle'' $\theta =
\arccos (-1/3)/2 \approx 54.74^{\circ}$.
Note that the additional delta-terms of the intra-trap dipolar interaction may be tuned to zero by means of a Feshbach resonance
~\cite{deuretzbacher2013erratum}. 
The effective interaction potential for particles in different traps does not have an equally simple form. With $n
\neq n^{\prime}$, we can turn the potential in
Eq.~(\ref{VDDIeff}) into a single integral expression,
\begin{align} 
& V^{\text{\it eff}}_{n,n'\ne n}(\tilde{x}) 
  =-\frac{{\bf d}^{2} } { 4\pi \gamma} \int_0^{\infty} dk k^2 e^{-k^2
    l^2_\bot/2-k \left|\tilde{x}\right|} \nonumber \\ & \times \left[
    \cos^2(\theta)\mathrm{J}_0(kz_w) +\cos^2(\phi)\mathrm{J}_2 (kz_w)
    \right. \nonumber \\ & - \left. \left( \frac{ \tilde{x}
      \sin(2\theta)\cos(\phi)}{\left|\tilde{x}\right|} +
    \frac{\sin^2(\theta)}{kz_w}\right)\mathrm{J}_1(kz_w) \right]
   , \label{VDDIinter}
  \end{align}
which has to be evaluated numerically. In Eq.~(\ref{VDDIinter}), 
$\mathrm{J}_i$ is the regular Bessel function of order $i$. 
At very large relative distances, however, the
interaction profile becomes similar to that of two particles in the same
trap, i.e. $\lim_{\lvert \tilde{x}\rvert \to  \infty} V^{\text{\it eff}}_{n,n'\ne n}(\tilde{x})
= V^{\text{\it eff}}_{n,n}(\tilde{x})$.

With the above assumptions, the  position representation of the effectively one-dimensional many-body Hamiltonian  for  $N_n$ 
particles in each trap $n$ reads 
\begin{align}
{H}^{\text{\it eff}} = & \sum_{n=0}^1 \left[ \sum_{i=1}^{N_n} {h}(x_{n,i}) + \sum_{j>i}^{N_n} V^{\text{\it eff}}_{n,n} \left( x_{n,i}-x_{n,j}
\right) \right] \nonumber \\  &+ \sum_{i=1}^{N_0}
\sum_{j=1}^{N_1} V^{\text{\it eff}}_{0,1} \left( x_{0,i} - x_{1,j}
\right) \label{Heff}
\end{align}
where $V^{\text{\it eff}}_{n,n^{\prime}}$ are given by Eqs.~(\ref{VDDIintra}, \ref{VDDIinter}) and where
\begin{equation}
{h}(x_{n,i}) = -\frac{\hbar^2}{2m}\frac{\partial^2}{\partial x^2_{n,i}} + \frac{m}{2}  \omega^2_x x^2_{n,i} \label{h}
\end{equation}
is the one-body operator associated with particle $i$ of trap $n$. Note
that the one-body confining potential $ m \omega^2_x x^2_{n,i}/2 =
V^{\text{trap}}_n\left(x_{n,i},0,nz_w\right)$. Also, we left out
the constant energy contribution $(N_0+N_1)\hbar \omega_\bot /2$,
associated with the combined oscillatory motion in the $yz$-plane,
from ${H}^{\text{\it eff}}$.

The eigenstates to the effective Hamilton operator,
Eq.~(\ref{Heff}), are obtained using the method of full configuration interaction, also known as exact diagonalization, in a $B$-spline basis; 
for further details see the appendix. 

\section{Ground-State Properties}

\begin{figure}[ht]
\includegraphics[width=0.9\linewidth]{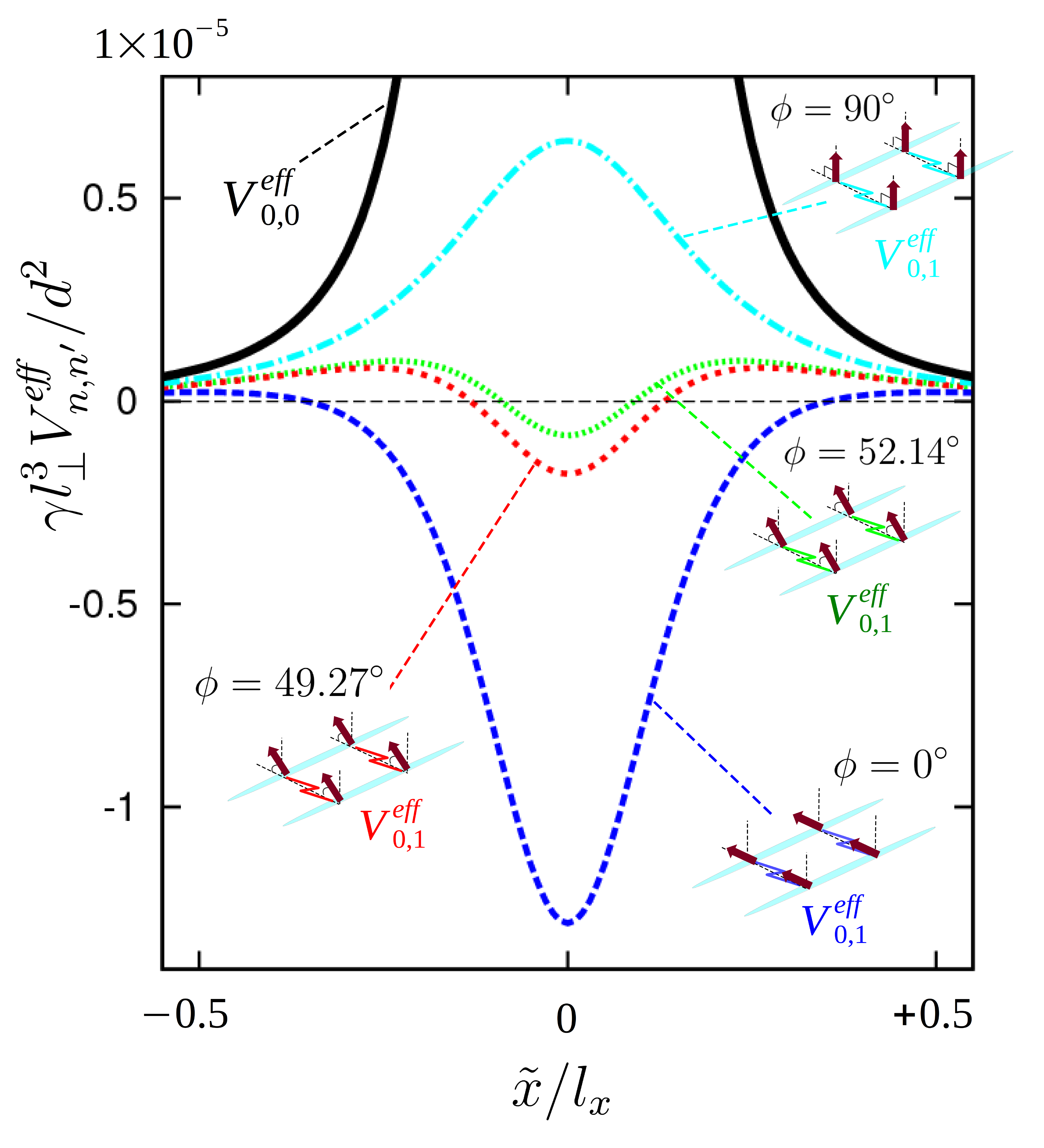}
\caption{(Color online) The effective intra- and inter-trap
  interaction potentials are shown for four different dipole
  alignments $(\theta,\phi) $ where $\theta  = 90^{\circ }$, for the parameters described in the main text. The intra-trap potential $V^{\text{\it eff}}_{0,0}$
  (black line) depends only on the angle $\theta$  and is thus identical in all cases, see Eq.~(\ref{VDDIintra}).  The inter-trap potential
  $V^{\text{\it eff}}_{0,1}$ (blue, red, green and cyan dashed lines)
  depends on both $\phi$ and $\theta$, see
  Eq.~(\ref{VDDIinter}).}
\label{fig2}
\end{figure}

We now proceed to investigate how  a change in  the 
relative dipole angle $\phi $ 
modifies the ground state properties of a few-body system.
To keep the numerical effort tractable, we restrict the total particle number to $N=4$, with $N_0=N_1=2$. These particles can be either 
spin-polarized fermions or spin-less bosons.  The traps are
characterized by the dimensionless parameters
$\omega_x/\omega_\bot = l^2_\perp/l_x^2= 10^{-4}$ , where $l_x=\sqrt{\hbar/m\omega_x}$, and $z_w/l_\bot = 25$. Clearly,
$\omega_x \ll \omega_\bot$ and $z_w > l_\bot$ which justifies the quasi
one-dimensional treatment of the system discussed in Sec.~\ref{sec:model}.
We choose an interaction strength dictated by the dimensionless
parameter $\mathbf{d}^2 m/ (\hbar^2 l_x \gamma )=
8\pi $ and an angle $\theta=90^{\circ}$ for a maximally repulsive 
{\it intra}-trap interaction. The  considered {\it intra}-trap potential $V_{n,n}^{\text {\it eff}}$
together with the {\it inter}-trap potentials $V_{n,n'\ne n}^{\text {\it eff}}$  obtained for different values of $\phi $ are shown in Fig.~\ref{fig2}. 

We here observe that the inter-trap potential has a short-ranged core, in terms of the $x$-coordinates for two particles in different traps, that can be made either repulsive or attractive depending on $\phi$, as well as a  repulsive long-ranged tail.

\begin{figure}[H]
\includegraphics[width=\linewidth]{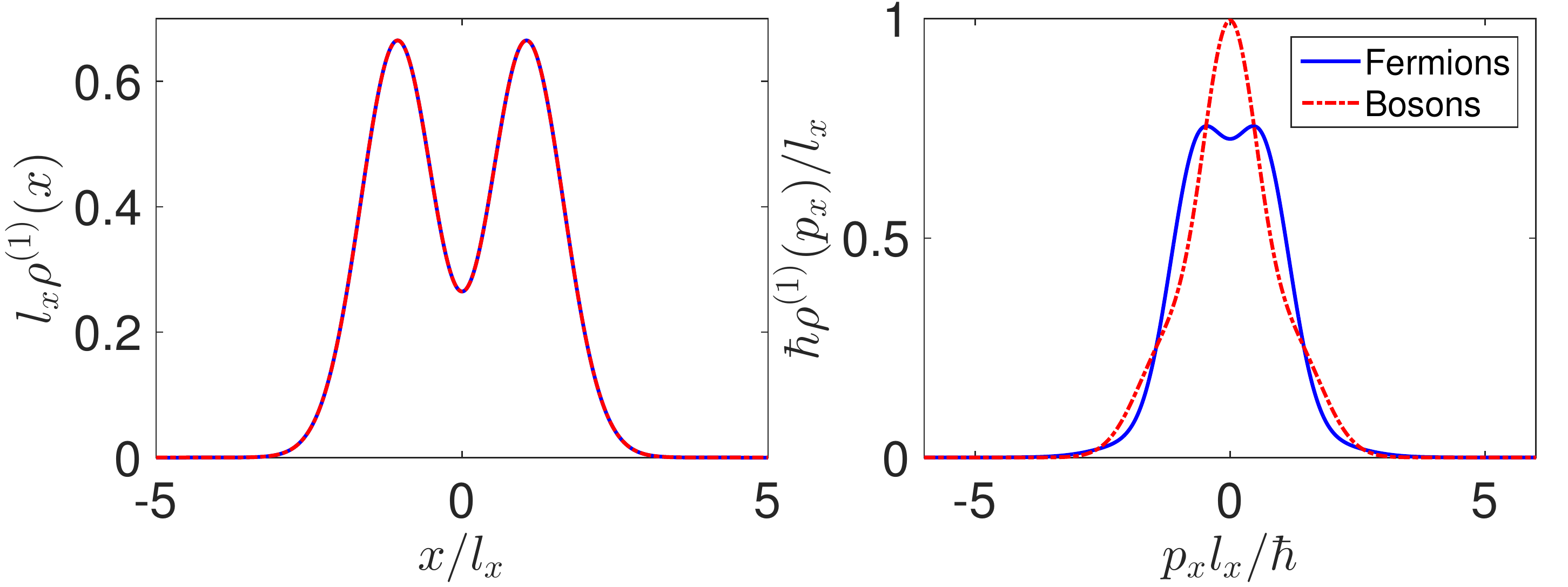} 
 \caption{ (Color online) The ground state density distributions
   computed for two particles in a single trap. 
   Spin-polarized fermions (solid blue line) are compared with spinless 
   bosons (red dashed line). {\it Left panel:} The
   single-particle density $\rho^{(1)}$ shows two pronounced peaks in
   the position space, being similar for fermions and bosons. {\it 
     Right panel:} The single-particle density in momentum space. The
   fact that the bosonic and fermionic distributions differ 
  agrees with the fact that  the ground state is not fully localized.}
\label{fig3}
\end{figure}

To begin with, let us ignore the interaction between particles
in different traps, thus effectively reducing the problem to that of a single trap with two particles.  Such a setup may be realized by moving the two wires infinitely far apart, i.e. $z_w\to\infty$. Figure~\ref{fig3} shows the single-particle densities for the ground state  in this particular case. We observe that the repulsive intra-trap interaction gives rise to similar single-particle density distributions $\rho ^{(1)}(x)$ for fermions and bosons. 
The pronounced minimum in $\rho (x)$ at $x=0$ reflects the onset of localization due to the long-range part of the DDI. 
In the limit of strong repulsion, $\rho^{(1)}(x=0)\rightarrow 0$ and a localized state is obtained, with identical   momentum distributions $\rho^{(1)}(p_x)$ for fermions and bosons, see~\cite{deuretzbacher2010}.
  
The question is now to what extent the inter-trap
interaction modifies the system properties.  For the considered case of 
$\theta=90^{\circ}$, 
one expects the densities to have common features with the 
ones in Fig.~\ref{fig3}, since $\left| V^{\text{\it eff}}_{n, n'\ne n}(\tilde {x})\right| < 
\left|V^{\text{\it eff}}_{n,n}(\tilde{x})\right|$. 
In particular, the strong repulsive intra-trap interaction leads to a partly
localized ground state.   
  
\begin{figure}[H]
\includegraphics[width=\linewidth]{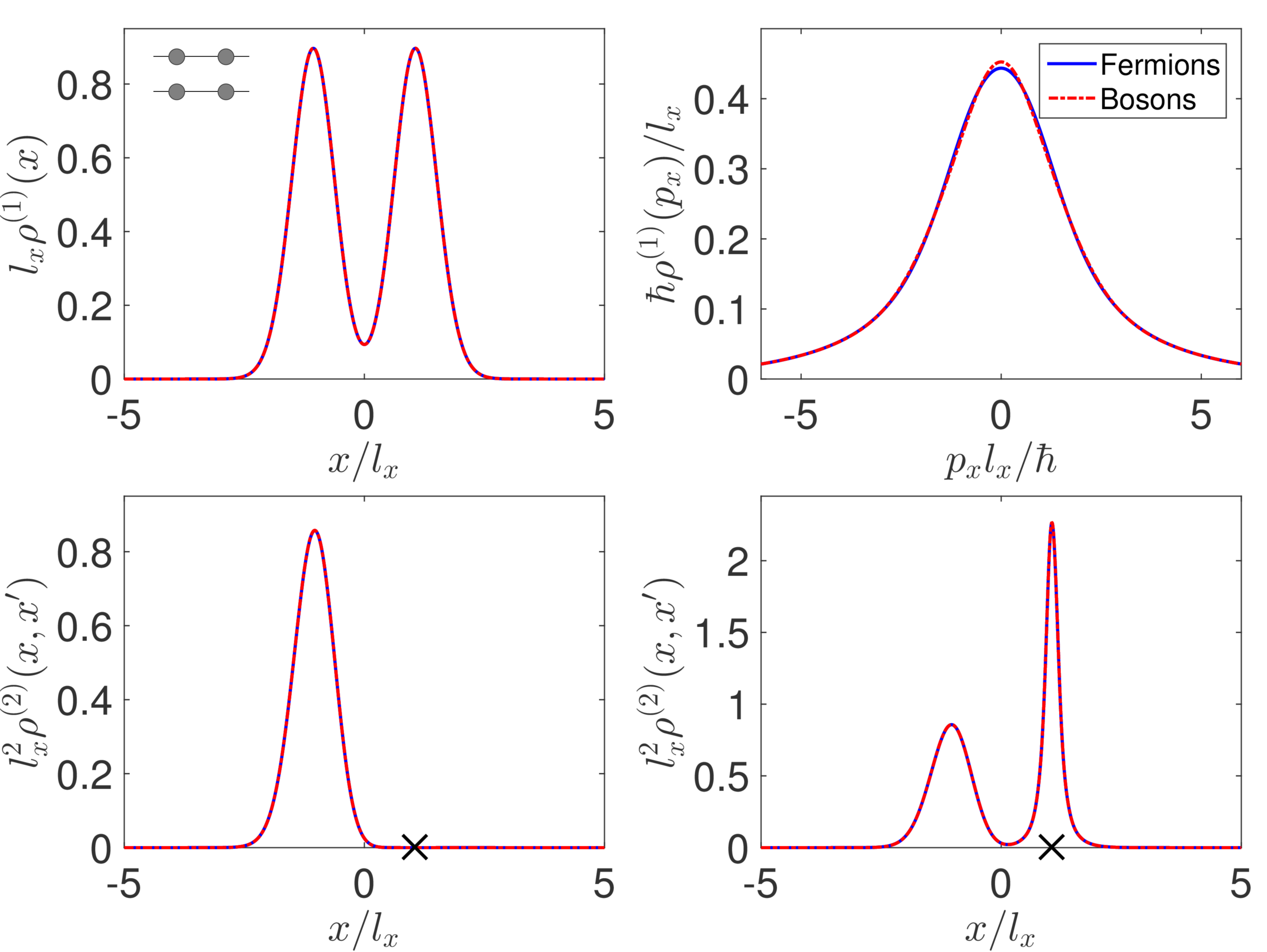} 
\caption{ (Color online) The ground state density distributions
within a single trap when $\phi=0.86 \approx 49.27^{\circ}$. 
The solid blue line is for fermions and the red dashed one for bosons. 
{\it Top left panel:} The single-particle density distribution clearly shows two pronounced peaks, indicating that the two particles in one of the traps are paired up with the corresponding ones in the other trap (see inset). 
{\it Top right panel:} The corresponding distribution in momentum space shows a similar behavior for fermions and bosons. Comparing to Fig.~\ref{fig3}, we note that the inter-trap interaction has further increased the localization. 
{\it Bottom left panel:} The pair-correlated density, with the position of the reference particle at $x^\prime$, indicated by the cross. Both reference particle and the considered particle reside in the same trap. 
{\it Bottom right panel:} The pair correlated density with the reference particle  in the other trap. The pronounced peak at $x=x^\prime $ indicates strong pairing between particles in different traps.
}
\label{FIG4}
\end{figure}

\begin{figure}[H]
\includegraphics[width=\linewidth]{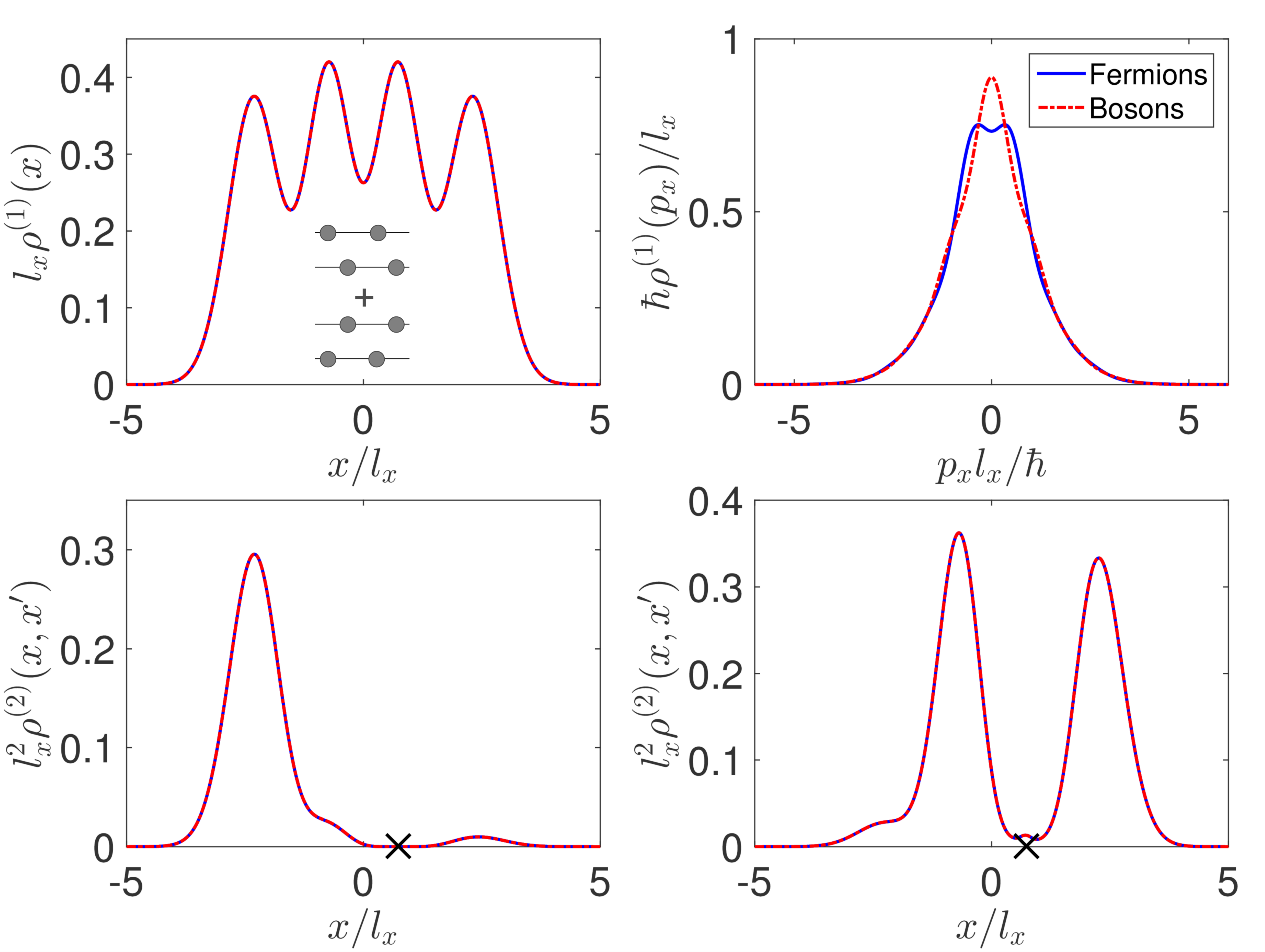} 
\caption{ (Color online) Similar to Fig.~\ref{FIG4}, but for $\phi=0.91 \approx 52.14^{\circ}$. The pattern of four distinct peaks in $\rho^{(1)}(x)$, top left panel, follows from the combination of a weaker attractive inter-trap core and a more pronounced repulsive tail, compared to that for lower values of $\phi$, making the interaction between particles in different traps effectively repulsive also at short relative distances. Hence, also the particles paired at lower $\phi$ will now repel one another, favoring separation in their relative position (see inset in the top left panel), as further supported by the pair-correlated densities in the lower two panels.}
\label{FIG5}
\end{figure}

In Fig.~\ref{FIG4} and~\ref{FIG5}, we show
the single trap density distribution at the two different angles $\phi = 0.86\approx 49.27^{\circ }$ and $\phi = 0.91\approx 52.14^{\circ }$. The intricate interplay between the attractive short-ranged core and the repulsive long-ranged tail of the inter-trap potential may, with such a minor change in $\phi$, significantly alter the properties of the system.  (The choice of values 
for the angle $\phi $ is bound by numerical limitations: For $\phi \lesssim 49^{\circ }$, the attractive core of the DDI becomes too strong to obtain accurate results with a tractable single-particle basis.)

With $\phi \approx 49.27^{\circ}$, shown in Fig.~\ref{FIG4}, the 
inter-trap interaction enhances the 
localization seen in the ground state densities.
Also, the bosonic and fermionic momentum distributions are now more similar than those in Fig.~\ref{fig3}. The localization is supported  both by the strong attractive core as well as the repulsive tail of the inter-trap interaction. The attractive part of the interaction leads to pairing between the particles in different traps at nearby $x$-coordinates, whereas the repulsive tail enhances the separation of {\it different} pairs of particles. Noteworthy is that the energy of the ground state is, for this particular value of $\phi$, greater than that of the corresponding system without inter-trap interaction. Hence, overall the inter-trap interaction is effectively repulsive. 
(In similar, but extended systems, a clustered crystal-like phase was predicted~\cite{knap2012}. The bound states found here appear as few-body pre-cursors to this cluster formation.)

Increasing the dipolar angle from $\phi\approx 49.27^{\circ }$ to $\phi\approx 52.14^{\circ }$, both the strength and range of the  inter-trap attractive core is reduced. In other words, particles in different traps start to repel when closer to one another in the $x$-direction .  We now observe four distinct peaks in the density profiles, see Fig.~\ref{FIG5}.
The four density peaks can be understood from the corresponding pair-correlated densities, clearly showing a particle displacement  in the single traps confining two particles each (see inset).
The two particles in each trap are now less localized than in Fig.~\ref{FIG4}, 
as indicated by the differing bosonic and fermionic momentum distributions. 

\begin{figure}[H]
\includegraphics[width=1.05\linewidth]{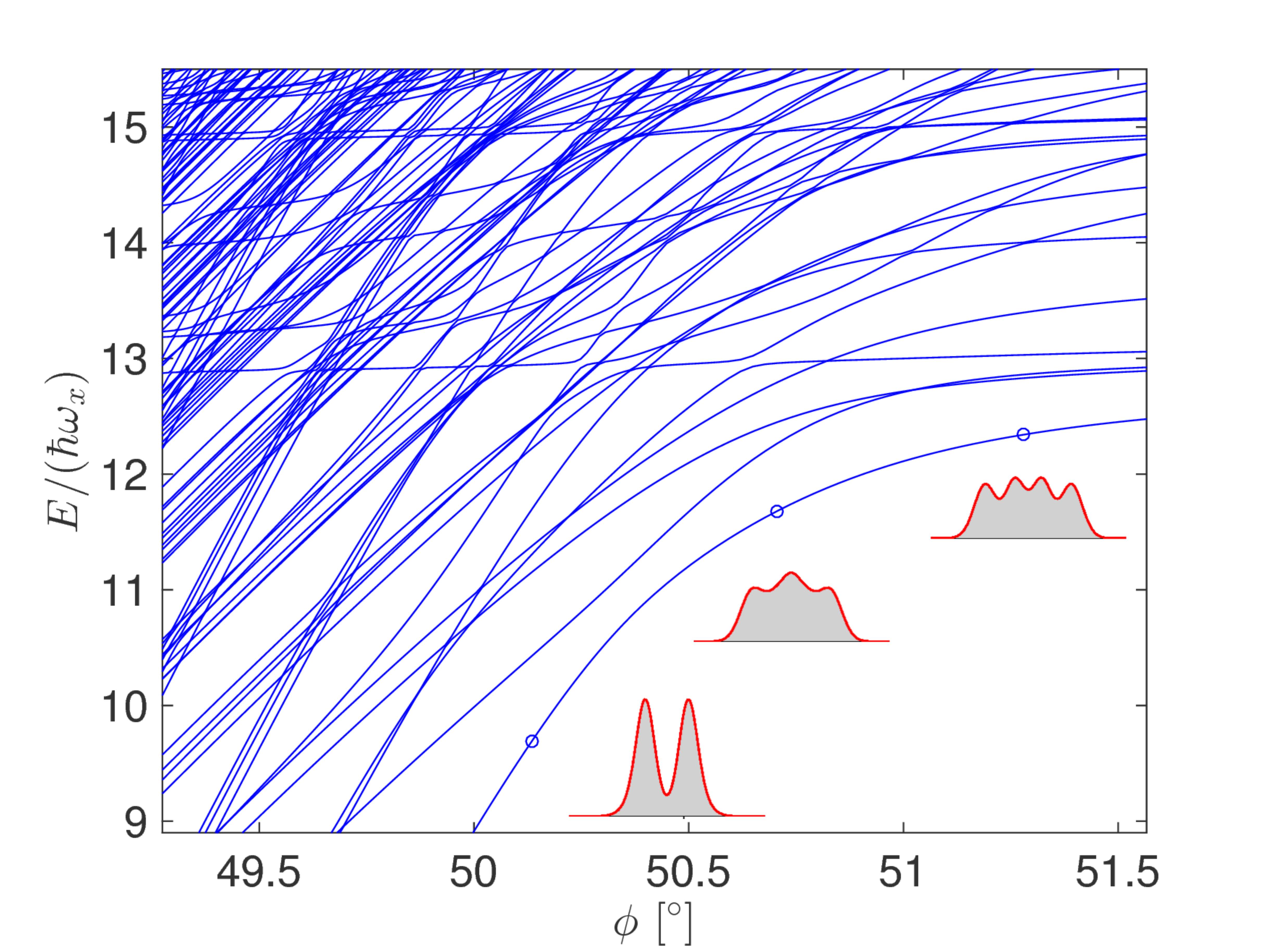}
 \caption{(Color online) The even parity spectrum for two spin-polarized fermions in each trap. The insets show the ground state density profiles for three different angles $\phi $ marked by circles in the main plot.}
\label{FIG6}
\end{figure}

\section{Quench of the dipole angle}

In Fig.~\ref{FIG6}, we show the
energy spectrum for even-parity states when two spin-polarized fermions are in each trap. (For comparison, without inter-trap interaction, the ground state energy of the corresponding system is $E \approx 5.66 \hbar\omega_x$). By decreasing the 
dipolar angle $\phi $, the ground state  seen in Fig.~\ref{FIG6} becomes more correlated and goes from non-localized to localized due to the pairing between particles in different traps
and its energy decreases. 
Interestingly, however, at higher energies in the spectrum,   
we note certain  energy levels that seem largely unaffected by the 
choice of $\phi$.  

The origin of these $\phi$-invariant energies can be understood by looking at the simpler system with a single particle in each trap, $N_0=N_1=1$.  In this case the Hamiltonian may be
separated into two parts: $H_R$ and $H_r$ describing the center of mass
motion of the two-body system and the relative motion of the two
particles, respectively. Here, the dipole-dipole interaction enters the
expression for $H_r$ alone,
\begin{equation}
H_r = \frac{p^2_{\tilde{x}}}{m} + \frac{1}{4}m\omega^2_{\bot}
\tilde{x}^2 + V_{n,n^{\prime}\neq n}^{\text{\it eff}}(\tilde{x}).
\end{equation}
Since the interaction potential is sharply peaked at $\tilde{x}=0$,
the odd parity solutions of $H_r$ are, with their nodes at this
position, largely unaffected by the nature of the dipole-dipole
interaction. The even parity solutions do, on the other hand, heavily
depend on the sign and strength of the interaction. Hence, when we
sweep $\phi$, we expect to see states described by odd parity
solutions in their relative coordinate manifested as interaction-invariant lines
in the spectrum. A similar structure was seen in the few-body spectra of contact-interacting bosons, where the few-body bound states depended strongly on the strength of the attractive interaction, while a second category of states appeared as interaction-invariant lines in the spectrum at strong attraction~\cite{tempfli2008excitations}. The interaction-invariant lines were identified as few-body precursors of fermionized super-Tonks-States~\cite{astrakharchik2004quasi,astrakharchik2005beyond}, where the short-range two-body correlations between the particles strongly affect he system properties. These similarities were also discussed in~Ref.~\cite{volosniev2013bound}
for dipolar interactions. 

\begin{figure}[h]
\includegraphics[width=0.8\linewidth]{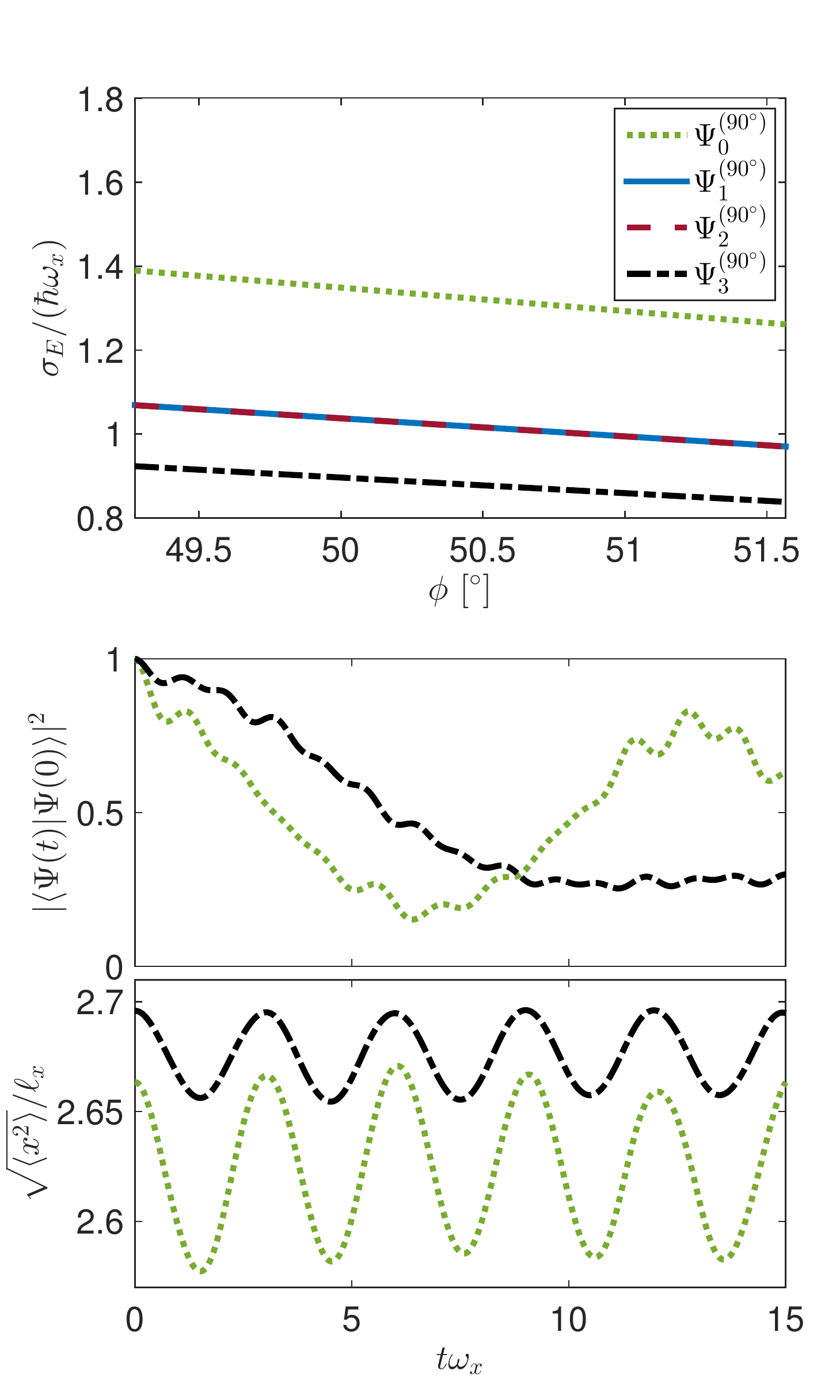}
\caption{(Color online) {\it Top panel:} Energy uncertainties $\sigma_E $, 
see Eq.~\ref{uncertainty}, as functions of the dipolar angle $\phi $. The plotted quantities represent the spread in energy after a fast
change of the dipolar angle from $90^\circ$ to $\phi$. Each line represents a different initial state of even parity before the quench. 
 {\it Middle panel:} The square modulus of the overlap between the initial state, before the quench, and the time-evolved one, after the quench at $t=0$. {\it Bottom panel:}  Square root of the expectation value of $x^2$ in a single-trap, illustrating the breathing mode dynamics of the system after the sudden quench. 
 (In the lower two panels, we consider only initial states $\Psi _0^{(90^{\circ})}$ and $\Psi _3^{(90^{\circ })}$ and a quench from  $\phi = 90^{\circ}$ to $\phi \approx 49.27^{\circ}$). 
 } \label{FIG7}
\end{figure}

Let us now investigate to what extent the lowest STG-like state,   see Fig.~\ref{FIG6}, can get populated. From the spectrum, we see that one possibility to reach this excited state is to start in a repulsive low-energetic eigenstate at a large $\phi$, i.e., $\phi=90^{\circ}$,  with $E\approx 13 \hbar \omega _x$, and then rapidly decrease the dipolar angle. 
Specifically, if the initial energy eigenstate 
remains an eigenstate to the Hamiltonian at a lower
$\phi$ value,  the desired single excited STG-like state is populated.  
Let us now check how closely the $i$:th energetically lowest state $\Psi_i^{(\phi=90^\circ)}$ relates to an eigenstate at lower $\phi$. For this
purpose, we compute 
\begin{equation}
\label{uncertainty}
\sigma_E = \sqrt{\langle \Psi_i^{(90^\circ)} \lvert H^{2}(\phi) \rvert \Psi_i^{(90^\circ)} \rangle -  \langle \Psi_i^{(90^\circ)} \lvert H(\phi) \rvert \Psi_i^{(90^\circ)} \rangle^2  }.
\end{equation}
If $\sigma_E=0$, then $\Psi_i^{(90^\circ)}$ is also an eigenstate to the
Hamiltonian $H(\phi)$ after the quench in $\phi$. In contrast,
$\sigma_E > 0$ means that non-stationary states are formed, and the
measure represents the shortest time scale that characterize a
significant change in the system, where the time scale decreases with
increasing $\sigma_E$.  In the top panel of Fig.~\ref{FIG7}, 
we plot $\sigma_E$ at
different $\phi$ for different initial states  of even parity.  
In the middle panel we clearly see that the third excited even-parity state 
at $\phi=90^\circ$, the one with lowest $\sigma _E$, is also closest to remain stationary after the quench. 
This observation agrees well with the fact that, at higher
values of $\phi$, the energy of the third excited state shown in
Fig.~\ref{FIG6} is largely unaffected by a change in the dipolar
angle. The non-zero contribution to $\sigma_E$ is mainly due to a small population of low-lying excited STG-like states, which depend only
weakly on $\phi$. A similar behavior has been seen in previous
attempts to populate ordinary STG states and gives rise to breathing
mode dynamics in the system~\cite{tschischik2015repulsive}, see the lowest panel of Fig.~\ref{FIG7}. In order to study the dynamics of the system after the quench, we here solve the time-dependent Schr\"{o}dinger equation numerically step-wise in time using a Krylov subspace~\cite{Park86}.
Recall that the small difference in $\sigma_E$ for the considered initial states implies that also systems prepared in the
ground state, as well as the first and second excited state, at
$\phi=90^\circ$ are close to stationary
after the quench. With a more strongly-peaked inter-trap  potential, the difference between the ground
state and third excited state at $\phi=90^\circ$ is expected to be
further reduced along with $\sigma _E$.

\section{Summary}

We studied the microscopic few-body properties of dipolar 
bosonic or fermionic particles confined in two parallel quasi-one-dimensional harmonic traps. 
By rotating the dipolar angle  $\phi$ about the trap axes and thereby changing the shape of the inter-trap dipole-dipole interaction, we found that the interplay between the short-  and long-range features of the inter-wire interaction had striking effects on the ground-state  properties of the system.

Starting in a weakly localized ground state at a relatively weak attractive core 
of the inter-trap interaction (weak in comparison to the repulsive tail),  a slow change in the dipolar angle brought the system into a regime of strong localization, owing to the formation of bound dimers between the traps.
Interestingly, a sudden quench in $\phi$ from a maximal repulsive core instead led to significant population-transfer into a class of excited states, analogous to super-Tonks-Girardeau states that are most commonly seen in strongly correlated one-dimensional systems. 
These states largely retain the structure of the non-localized ground state at $\phi=90^\circ$. For a sudden quench in the dipolar angle the localization of particles is therefore inhibited.

\section*{ACKNOWLEDGMENTS}

We  thank P. D'Amico, M. Rontani and J. Cremon for valuable discussions and input at an initial stage of our work. The research was financially supported by the Swedish Research Council and NanoLund at Lund University.   

\vfill\eject

\section*{Appendix: B-splines and exact diagonalization}

$B$--splines
are piece-wise polynomials that frequently are defined through the
recursive relation \cite{de1978practical}
\begin{align}
B_{i,1}(x) & = \left\{ \begin{array}{ll}
1 & \text{ if } \tau_i \leq x \leq \tau_{i+1}\\
0 & \text{ otherwise }\end{array} \right. \\
B_{i,k}(x) & =  \frac{x-\tau _i}{\tau_{i+k-1}-\tau_i}B_{i,k-1}(x) + \frac{\tau_{i+k}-x}{\tau_{i+k}-\tau_{i+1}}B_{i+1,k-1}(x)
\end{align}
where $\tau_i \geq \tau_{i+i} $ are the so-called knot-points and $k$ is the
order of the $B$-splines. Throughout this work we set $k=5$. We use a linear central distribution
of knot-points with $\Delta \tau=0.12$, and exponentially increasing distances between the knot-points outside of the central region. The outermost knot-points were placed
at $x=\pm 5$. In total we used 64 $B$-splines (except for the spectra shown in Fig.~\ref{FIG6} and the lower panels of Fig.~\ref{FIG7}, where the number of $B$-splines was 56 to reduce the computational workload). 

First, we construct an orthonormal one-body basis for the $\varphi _j$ in 
Eq.~\ref{one-body orb} by diagonalization of the Hermitian $h$,  Eq.~(\ref{h}), in the $B$-spline basis. We write
\begin{equation}
\varphi_j(x) = \sum_ic_{j,i}B_{k,i}(x)
\end{equation}
which turns the one-body Schr\"{o}dinger equation, 
$h(x)\varphi_j(x)= \epsilon_j \varphi_j(x)$, into the generalized eigenvalue problem,
\begin{equation}
\mathbf{h}\mathbf{c}_j = \epsilon_j \mathbf{S}\mathbf{c}_j  
\end{equation}
with matrix elements given by
\begin{align}
\mathbf{h}_{i,j} &= \int dx B_i(x)h(x)B_j(x),\label{hm}\\
\mathbf{S}_{i,j} &= \int dx B_i(x)B_j(x).\label{sm}
\end{align}

 Finally, an orthonormal many-body basis is constructed
based on correctly symmetrized products of the acquired (orthogonal)
one-body states following the general prescription of the
configuration interaction method. 


\end{document}